\documentclass[10pt, conference]{IEEEtran}
\IEEEoverridecommandlockouts
\usepackage[noadjust]{cite}
\usepackage{amsmath, amssymb, amsfonts, url}
\usepackage[ruled,vlined]{algorithm2e}
\usepackage{graphicx, tabularx, booktabs, subcaption, multirow}
\usepackage{textcomp}
\usepackage{xcolor}
\usepackage[acronym]{glossaries}

\DeclareMathOperator*{\argmin}{arg\min}
\def\BibTeX{{\rm B\kern-.05em{\sc i\kern-.025em b}\kern-.08em
    T\kern-.1667em\lower.7ex\hbox{E}\kern-.125emX}}

\begin{document}

\title{Efficient DCT-Based Estimation and Compensation of Nonlinear Channels for OFDM Systems
\thanks{This work is part of the project SOFIA PID2023-147305OB-C32 funded by MICIU/AEI/10.13039/501100011033 and FEDER/UE.}
}

\author{\IEEEauthorblockN{Marc Martinez-Gost\IEEEauthorrefmark{1}, Ana Pérez-Neira\IEEEauthorrefmark{1}\IEEEauthorrefmark{2}\IEEEauthorrefmark{3}, Miguel Ángel Lagunas\IEEEauthorrefmark{2}}
\IEEEauthorblockA{
\IEEEauthorrefmark{1}Centre Tecnològic de Telecomunicacions de Catalunya, Spain\\
\IEEEauthorrefmark{2}Dept. of Signal Theory and Communications, Universitat Politècnica de Catalunya, Spain\\
\IEEEauthorrefmark{3}ICREA Acadèmia, Spain\\
Corresponding author: mmartinez@cttc.es
}}

\newacronym{AI}{AI}{Artificial Intelligence}
\newacronym{AirComp}{AirComp}{over-the-air computation}
\newacronym{AWGN}{AWGN}{additive white Gaussian noise}
\newacronym{CNN}{CNN}{convolutional neural network}
\newacronym{CSI}{CSI}{channel state information}
\newacronym{DA}{DA}{direct aggregation}
\newacronym{DSB}{DSB}{double sideband}
\newacronym{FL}{FEEL}{federated learning}
\newacronym{FSK}{FSK}{frequency shift keying}
\newacronym{MSE}{MSE}{mean squared error}
\newacronym{NMSE}{NMSE}{normalized mean squared error}
\newacronym{PAM}{PAM}{pulse amplitude modulation}
\newacronym{PPM}{PPM}{pulse position modulation}
\newacronym{TBMA}{TBMA}{type-based multiple access}
\newacronym{SNR}{SNR}{signal-to-noise ratio}

\maketitle
\begin{abstract}
This paper proposes a maximum-likelihood (ML) framework for estimating nonlinear frequency-selective channels in  orthogonal frequency-division multiplexing (OFDM) communication systems. The nonlinear distortions are modeled using a Discrete Cosine Transform (DCT)-based representation, which results in a well-conditioned estimation problem with favorable convergence properties. The proposed method combines a compact parameterization with low computational complexity, enabling fast adaptation and efficient real-time implementation. Numerical results show that the proposed channel estimation can be used for multiple nonlinear compensation methods and achieve near-ideal BER performance with very limited training overhead. This performance is maintained down to approximately 15 dB SNR in the presence of both amplitude and phase nonlinearities, and down to -10 dB when only amplitude distortions are present. 

\end{abstract}

\section{Introduction}

Orthogonal frequency-division multiplexing (OFDM) has become the dominant waveform in modern wireless communication systems due to its high spectral efficiency, robustness against multipath propagation, and compatibility with high-order modulation schemes. However, OFDM signals inherently exhibit a high peak-to-average power ratio (PAPR), producing large signal peaks that can push the power amplifier (PA) into its nonlinear operating region \cite{papr_review}. Preventing this distortion requires operating the PA with a substantial input back-off, which significantly reduces its power efficiency. This trade-off is particularly critical in power-constrained systems such as satellite communications \cite{HPA_sat}, where efficient operation of the onboard PA is essential to maximize the available transmit power. Therefore, compensating for PA nonlinearities remains an important challenge in the design of efficient OFDM communication systems.

Digital predistortion (DPD) has long been the predominant approach for mitigating PA nonlinearities by linearizing the transmitter before amplification \cite{PA_modeling}. Its effectiveness, however, relies on accurately identifying the PA nonlinear response, a task that has become increasingly challenging in modern communication systems \cite{lopez_bueno_2023}. The wider bandwidths and stronger nonlinear behavior encountered in current standards demand higher sampling rates and increasingly sophisticated behavioral models, substantially increasing computational complexity. Furthermore, the adoption of flexible physical-layer configurations requires adaptive DPD schemes capable of continuously tracking PA variations, which introduces additional training overhead. As a result, closed-loop PA estimation at the transmitter side becomes increasingly demanding in terms of power consumption.

Digital postdistortion (DPoD) has emerged as an attractive alternative to DPD \cite{post-distortion}. DPoD performs the compensation at the receiver, potentially accounting for impairments accumulated along the transmission chain. The effectiveness of DPoD, however, still depends on accurate estimation of the nonlinear channel. Existing approaches often assume that the multipath channel is known or that it can be estimated independently of nonlinear distortions, an assumption that is difficult to satisfy in practical systems (see \cite{post-distortion} and references therein). Furthermore, existing methods do not provide a unified framework capable of jointly estimating the multipath channel response together with amplitude and phase distortions.

More recently, data-driven approaches based on neural networks have been proposed to avoid explicit modeling and estimation of the nonlinear channel \cite{nn_heavy,11104481,dl_geoffrey}. Although these methods claim considerable modeling flexibility, their receiver-side nonlinear processing introduces additional challenges. In particular, nonlinear transformations of the received signal alter the statistical properties of the noise, potentially amplifying its impact and making the learning process more difficult. Consequently, neural network-based DPoD approaches typically require large model architectures and extensive training datasets to achieve satisfactory performance. This high training complexity significantly limits their applicability in adaptive scenarios where the nonlinear channel must be continuously tracked.

These limitations are expected to become increasingly relevant as future wireless systems place greater emphasis on uplink communications. Accurate knowledge of the nonlinear channel at the user equipment (UE) enables more efficient transmission strategies. Since UEs operate under stringent computational and energy constraints, shifting the nonlinear estimation task to the receiver provides an attractive alternative to conventional transmitter-centric approaches. Moreover, receiver-side estimation naturally captures the aggregate nonlinear behavior of the entire transmission chain, rather than only the PA characteristics, making it applicable to a broader range of practical impairments.

Motivated by these challenges, this work proposes a unified framework for nonlinear frequency-selective channel estimation for OFDM systems. The proposed formulation is fully consistent with the maximum-likelihood (ML) principle and models amplitude and phase nonlinear distortions using a Discrete Cosine Transform (DCT) model. Unlike conventional polynomial models, the cosine basis of the DCT fundamentally improve the conditioning of the estimation problem by effectively decorrelating the optimization variables and minimizing the eigenvalue spread of the parameter space \cite{dct_apn}. This property enables efficient gradient-based optimization with low computational complexity and fast convergence. In particular, the resulting algorithm is well suited to real-time operation in time-varying scenarios where channel statistics are unknown or continuously evolving.


The contributions of this work\footnote{This article is an extended version of a previously published conference paper \cite{martinez2026dct}, in which only amplitude distortions are considered.} are as follows:
\begin{enumerate}
    \item We propose an ML–based estimation framework for nonlinear frequency-selective channels, jointly capturing multipath, amplitude and phase distortions.

    \item The nonlinear channel is represented with a DCT-based model, which provides a compact and flexible parameterization of the nonlinear responses. The favorable spectral properties of the DCT lead to fast and well-conditioned convergence. In addition, the formulation does not require step-size tuning or the estimation of second-order statistics. Consequently, the proposed framework enables fast adaptation with low computational complexity, making it particularly well suited for real-time operation.

    \item We show how the proposed estimation framework can be integrated into different compensation strategies. In particular, for predistortion, the DCT structure can be exploited to efficiently approximate the inverse nonlinear response with low computational complexity.
\end{enumerate}

The remaining part of the paper proceeds as follows: Section II presents the system model, followed by Section III, which proposes the nonlinear channel estimation design. Section IV then describes the compensation techniques and discusses their integration with the channel estimation framework. Section V presents the performance evaluation and Section VI concludes the paper.

\textit{Notation:} Bold lowercase and uppercase letters denote vectors and matrices, respectively. The operators $(\cdot)^T$, $(\cdot)^*$, and $(\cdot)^H$ denote transpose, complex conjugate, and Hermitian transpose, respectively. The sets of real and complex numbers are denoted by $\mathbb{R}$ and $\mathbb{C}$. The operator $\operatorname{diag}(\cdot)$ forms a diagonal matrix from its argument and $\mathbf{I}$ is the identity matrix. The expectation operator is denoted by $\mathbb{E}(\cdot)$. The real part, imaginary part and phase of a complex scalar are denoted by $\Re({\cdot})$, $\Im({\cdot})$, and $\angle(\cdot)$, respectively. The notation $\mathcal{O}(\cdot)$ denotes the asymptotic order in the Big-O sense.

\section{System model}
\begin{figure}[t]
    \centering
    \includegraphics[width=\columnwidth]{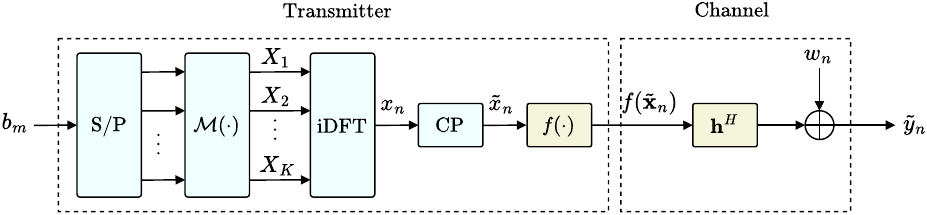}
  \caption{System model of the OFDM transmission chain, including signal generation, nonlinear distortion, multipath channel propagation and additive noise.}
  \label{fig:rx_estimation}
\end{figure}

This section describes the complete OFDM transmission model, including signal generation and nonlinear frequency-selective distortion. We further introduce a DCT-based parametrization of the nonlinear amplitude and phase distortions, which is used throughout the remainder of the work for modeling and estimation. The overall system model is illustrated in Figure~\ref{fig:rx_estimation}.

\subsection{OFDM transmitted signal}
In OFDM, an incoming serial stream of bits $\{b_1,\dots, b_M\}$ is  parallelized into multiple streams and each one mapped to a complex symbol stream using some modulation constellation $\mathcal{M}(\cdot)$. Each symbol $X_k\in\mathbb{C}$ is mapped to an orthogonal subcarrier signal and transmitted simultaneously. This procedure is realized with an inverse discrete Fourier transform (iDFT) and results in the so-called OFDM symbol:
\begin{equation}
    x_n = \mathcal{F}^{-1}\left(X_k\right)
    =\sum_{k=0}^{K-1} X_k \text{e}^{j\frac{2\pi kn}{K}}, \quad n=0,\dots,K-1,
    \label{eq:idft}
\end{equation}
where $K$ is the resolution of the DFT and determines the number of subcarriers. 

Typically, a cyclic prefix of length $\text{CP}\geq L-1$ is incorporated by appending the last CP samples of \eqref{eq:idft} at the beginning of the signal to combat intersymbol interference (ISI) and simplify channel estimation. We can express the resulting signal as follows:
\begin{equation}
    \tilde{x}_n =
    \begin{cases}
    x_{n + N_\text{DFT} - \text{CP}}, & n = 0,\dots,\text{CP}-1, \\[6pt]
    x_{n-\text{CP}}, & n = \text{CP},\dots,N_\text{DFT}+\text{CP}-1
    \end{cases}
    \label{eq:ofdm_symbol}
\end{equation}


The OFDM signal $\tilde{x}_n$ passes through a memoryless nonlinear PA, whose behavior is characterized by amplitude-to-amplitude (AM-AM) and amplitude-to-phase (AM-PM) distortions. We will also refer to these nonlinearities as magnitude and phase distortions. The resulting output signal can be expressed as
\begin{align}
    f(x_n)=f_\text{AM}(|x_n|)\,\text{e}^{j\left(
    \phi_n+f_\text{PM}(|x_n|)
    \right)},
    \label{eq:NLD}
\end{align}
where $f_\text{AM}(\cdot)$ and $f_\text{PM}(\cdot)$ denote the AM-AM and AM-PM characteristics of the PA, respectively, and $\phi_n=\angle\,x_n$. This formulation encompasses commonly used memoryless PA models, such as the Saleh, Rapp and Ghorbani models \cite{working_group}, where the amplitude of the signal dictates the distortions at the output of the amplifier. We assume that the AM-AM characteristic is normalized such that $0 \le |f_\text{AM}(x_n)| \le 1$.

\subsection{Nonlinear Channel}

The signal is then transmitted through a frequency-selective channel with multipath propagation. The channel is modeled as a finite impulse response (FIR) filter $\mathbf{h}\in\mathbb{C}^L$, where $L$ denotes the number of taps. The channel coefficients are assumed to be i.i.d. complex Gaussian, $h_\ell \sim \mathcal{CN}(0,1/L)$, ensuring unit-normalized channel power. Throughout this work, we assume a block-fading channel, so that the channel remains constant over the duration of a processing block.

The received signal is given by
\begin{equation}
    \tilde{y}_n =\mathbf{h}^H f(\tilde{\mathbf{x}}_n)+w_n=\sum_{\ell=0}^{L-1}h_\ell^*f(\tilde{x}_{n-\ell})+w_n,
    \label{eq: rx_ofdm}
\end{equation}
where $\mathbf{x}_n=[x_n,x_{n-1},\dots,x_{n-L}]^T$ collects $L$ consecutive transmitted samples, the nonlinearity is applied element-wise, and $w_n$ is circularly symmetric complex Gaussian noise with variance $\sigma^2$.

Assuming that the cyclic prefix length satisfies $\mathrm{CP}\ge L-1$, the receiver removes the cyclic prefix and obtains the useful OFDM symbol
\begin{align}
    y_n 
    &= \sum_{\ell=0}^{L-1} h_\ell^* 
       f\left({x}_{[n-\ell\text{ mod } N_\text{DFT}]}\right) 
       + w_{n+\text{CP}}, 
    \label{eq:rx_cp_ofdm}
\end{align}
where $[(\cdot)\text{ mod }{N_\mathrm{DFT}}]$ denotes the modulo-$N_\mathrm{DFT}$ operation. Under this condition, the linear channel convolution is transformed into an equivalent circular convolution over the useful OFDM symbol interval.

For the derivations that follow, we restrict the analysis to the samples $n=L-1,\dots,N_\mathrm{DFT}-1$. Since these indices do not involve wrap-around, the modulo operation can be omitted and the received signal can be expressed using the equivalent linear-convolution notation.

The receiver cannot distinguish between the phase rotation introduced by the channel and that introduced by the nonlinear distortion, since both contributions appear additively in the received phase. To resolve this ambiguity, we assume without loss of generality that the first channel tap has zero phase (i.e., $h_0\in\mathbb{R}$) and absorb the corresponding phase rotation into the nonlinear phase distortion model. Although this ambiguity cannot be resolved without additional information, it does not affect the receiver operation, whose ultimate objective is to compensate the overall phase distortion introduced by the nonlinear channel.

The considered system model corresponds to a Hammerstein structure, in which an instantaneous nonlinear block is followed by a linear FIR filter \cite{PA_modeling}. A more general formulation capable of describing nonlinearities with memory would require a Wiener-Hammerstein model, where a linear FIR filter precedes the instantaneous nonlinearity. Such an extension would allow the representation of dynamic nonlinear effects and memory-dependent distortions. However, the incorporation of memory effects into the nonlinear model is beyond the scope of this work and is left for future research.



\subsection{The DCT-based model for nonlinearities}
The receiver needs to model the nonlinear distortions in \eqref{eq:NLD} with a parametrized model. In this work we propose to use the DCT \cite{dct_apn}. For a nonlinear function $\hat{f}(\cdot)$, its DCT representation is
\begin{equation}
    \hat{f}(z)=\sum_{q=1}^{Q}F_q\cos\left(\frac{\pi}{2N_\text{DCT}}(2q-1)(2z+1)\right),
    \label{eq:dct}
\end{equation}
where $F_q\in\mathbb{R}$ are the DCT coefficients, $Q$ is the total number of coefficients, $z\in[0,N_{\text{DCT}}-1]$ is a discrete index and $N_{\text{DCT}}$ is the number of discrete samples that defines the resolution of the DCT. Although the DCT is defined over a discrete input, it can represent any input signal because the cosine basis functions provide interpolation between those points. Thus, keeping $N_\text{DCT}$ large enough (e.g., 512) is sufficient and the DCT has no hyperparameters beyond $Q$.

Since the nonlinear model is defined over a normalized input magnitude domain $|x_n| \in [0,1]$, a linear mapping is introduced to project this interval onto the discrete DCT index domain:
\begin{equation}
    z_n = (N_\text{DCT}-1)\frac{x_n+1}{2}
\end{equation}

This transformation ensures compatibility between the continuous-valued signal magnitude and the discrete grid required by the DCT representation. Moreover, this mapping effectively extends the domain to $[-1,1] \mapsto [0,N_{\text{DCT}}-1]$, which is a deliberate modeling choice used to enforce an odd-symmetric structure in the approximated nonlinear function around $x_n=0$. Although the input magnitude is inherently nonnegative, this symmetry assumption is introduced to guarantee that the nonlinear response satisfies $\hat{f}(x_n=0)=0$, which reflects the natural constraint that no distortion should be present in the absence of an input signal. Additionally, this structural property enables a sparse parameterization of the model, since only odd-indexed DCT coefficients contribute to the representation. This is enforced by the term $(2q-1)$ in \eqref{eq:dct}, which significantly reduces the number of parameters required to describe the nonlinear function. 

Finally, equation \eqref{eq:dct} can be expressed as $\hat{f}(z_n)=\mathbf{f}^T\mathbf{c}_n$, where $\mathbf{f}=[F_1,F_2,\dots,F_Q]^T\in\mathbb{R}^Q$ contains the $Q$ DCT coefficients and $\mathbf{c}_n\in\mathbb{R}^Q$ contains the cosines evaluated at $z_n$. We can generalize this notation when the input is a vector $\mathbf{x}_n$ to $\hat{f}(\mathbf{x}_n)=\mathbf{C}_n^T\mathbf{f}$, where
\begin{equation}
    \mathbf{C}_n=\left[\mathbf{c}_{n},\mathbf{c}_{n-1},\dots, \mathbf{c}_{n-\ell},\dots, \mathbf{c}_{n-(L-1)}\right]\in\mathbb{R}^{Q\times L}
\end{equation}


As introduced in \cite{dct_apn}, the output $\hat{f}(x_n)$ can be compared with a reference signal $f(x_n)$ to generate an error signal and update the DCT coefficients. In this work we will use the DCT to model the distortions estimates $\hat{f}_\text{AM}(\cdot)$ and $\hat{f}_\text{PM}(\cdot)$ at the receiver side, and define the corresponding error signals required to update the coefficients.

The proposed DCT-based model extends the conventional Least mean squares (LMS) framework by introducing nonlinearity through the basis expansion while preserving linear adaptation with respect to the filter coefficients. Owing to the orthogonality and boundedness of the DCT kernels, the resulting input correlation matrix becomes diagonal and independent of the input statistics, $\mathbf{R}_c=\mathbb{E}\left\{\mathbf{c}_n \mathbf{c}_n^T\right\}=\frac{1}{2}\mathbf{I}$. This leads to simplified step-size selection, predictable convergence behavior and minimum eigenvalue spread. As a result, the proposed approach provides an efficient and theoretically well-conditioned framework for adaptive estimation of nonlinear distortions \cite{dct_apn}. The following section shows how this framework can be integrated into the considered OFDM receiver to jointly estimate the linear and nonlinear distortions introduced by the channel.

\section{Nonlinear Channel Estimation}
\label{sec:estimation}

In conventional multi-carrier receivers, the channel response is typically estimated in the frequency domain, where the channel decomposes into $K$ narrowband subchannel responses. However, when the transmitted signal undergoes magnitude distortion, channel estimation becomes more challenging because the nonlinear distortion occurs in the time domain and it creates inter-carrier interference (ICI) in the frequency domain. Moreover, there is no analytical expression describing its effect on the distorted signal in the frequency domain. 

To address this limitation, we propose to estimate the full effective nonlinear channel response in the time domain. Thus, the whole OFDM symbol (i.e., the full sequence of bits $\{b_1,\dots,m_M\}$) is treated as a pilot observation for joint channel and nonlinearity estimation.

Next, we first formalize the estimation problem and introduce the optimal design criterion. Because the resulting optimization problem is nonconvex, we develop a block coordinate descent algorithm and leverage the DCT properties to enable efficient computation and fast convergence.

\subsection{Optimal Design Criterion}

An optimal ML receiver operates by first characterizing the communication channel, in order to model the transformations experienced by the transmitted signal. Then, the receiver should select the most probable $\hat{\mathbf x}_n$ given $y_n$. Under additive white Gaussian noise (AWGN) and complete knowledge of the nonlinear channel response, the conditional probability density function of the received signal is
\begin{equation}
    p\left(y_n \mid \mathbf{h}^H{f}(\mathbf{x}_n)\right) = \frac{1}{{\pi\sigma^2}}\, \exp\left( -\frac{\big|y_n - \mathbf{h}^H{f}(\mathbf{x}_n)\big|^2}{\sigma^2} \right)
\end{equation}
and $\hat{\mathbf x}_n$ is obtained maximizing the log-likelihood, $\log p\left(y_n \mid \mathbf{h}^H{f}(\mathbf{x}_n)\right)$. This results in
\begin{equation}
    \hat{x}_n = \argmin_{} \, \big|y_n - \mathbf{h}^H{f}(\mathbf{x}_n)\big|^2,
    \label{eq:optimal_ml_channel}
\end{equation}
revealing that the ML solution corresponds to the $\mathbf x_n$ that minimizes the mean squared error (MSE) between the observed data $y_n$ and the signal replica $\mathbf{h}^H{f}(\mathbf{x}_n)$. Therefore, the optimal ML estimator must replicate the Hammerstein model. 

Given a pilot sequence $\mathbf{x}_n$ and estimates of both the channel taps, $\hat{\mathbf{h}}$, and the nonlinear distortion, $\hat{f}(\cdot)$, the receiver computes an estimate of the received signal,
\begin{align}
    \hat{y}_n=\hat{\mathbf{h}}^H \hat{f}(\mathbf{x}_n)=
    \sum_{\ell=0}^{L-1}\hat{h}_\ell^*\hat{f}_\text{AM}(|x_{n-\ell}|)
    \,\text{e}^{j\left(\phi_{n-\ell}+\hat{f}_\text{PM}(|x_{n-\ell}|)\right)}
\end{align}

This signal can be expressed in terms of the nonlinear DCT model as
\begin{align}
    \hat y_n
    &=
    \sum_{\ell=0}^{L-1}
    \hat h_\ell^*\,
    \mathbf f_\mathrm{AM}^T \mathbf c_{n-\ell}\,
    e^{j\left(
        \phi_{n-\ell}
        +\mathbf f_\mathrm{PM}^T\mathbf c_{n-\ell}
    \right)}
    \nonumber\\
    &=
    \hat{\mathbf h}^H \mathbf D_n
    \mathbf C_n^T\mathbf f_\mathrm{AM},
    \label{eq:rx_estimate}
\end{align}
where $\mathbf{f}_\text{AM}$ and $\mathbf{f}_\text{PM}$ represent the coefficients of the magnitude and phase distortions, respectively. Matrix 
$\mathbf{D}_n=\operatorname{diag}\!\left(
        e^{j\left(
            \boldsymbol{\phi}_n
            +\mathbf C_n^T\mathbf f_\mathrm{PM}
        \right)}
    \right)\in\mathbb{C}^{L\times L}$, $\boldsymbol{\phi}_n=\left[\phi_n, \phi_{n-1},\dots, \phi_{n-(L-1)}\right]^T$ contains the phases of the original reference signal, and the complex exponential is applied element-wise.

Finally, the receiver formulates the instantaneous error by comparing the observed and estimated received signals,
\begin{equation}
    \varepsilon_n=y_n-\hat{y}_n=y_n-\hat{\mathbf h}^H \mathbf{D}_n
    \mathbf C_n^T\mathbf f_\mathrm{AM},
    \label{eq:error}
\end{equation}
and the nonlinear channel is estimated by minimizing the MSE,
\begin{align}
    \operatorname*{minimize}_{\hat{\mathbf{h}},\, \mathbf f_{\mathrm{AM}},\,\mathbf f_{\mathrm{PM}}}
    \quad
    & \mathbb{E}\left\{|\varepsilon_n|^2\right\},
    \tag{P1}
    \label{eq:min_mse}
\end{align}
where the expectation is taken with respect to the transmitted data and the channel noise realization.

Nevertheless, the optimization problem in \eqref{eq:min_mse} is not jointly convex, because the estimated signal $\hat{y}_n$ contains multiplicative couplings between the channel coefficients and the magnitude DCT coefficients, and the phase distortion parameters appear inside a complex exponential. This renders the MSE nonconvex, which precludes the use of standard convex optimization tools to obtain a closed-form global solution.

To address this issue, we adopt a block-coordinate descent strategy, where one set of parameters is optimized at a time while the remaining variables are held fixed. In particular, the optimization alternates among the channel taps $\hat{\mathbf h}$, the magnitude DCT coefficients $\mathbf f_{\text{AM}}$, and the phase DCT coefficients $\mathbf f_{\text{PM}}$. In the following, we study each subproblem individually, examining its convexity properties. We ultimately develop an efficient iterative algorithm to compute a stationary point of the original nonconvex formulation.


\subsection{Frequency-selective Channel}

This section focuses on the estimation of the channel taps $\mathbf{h}$. We assume that the receiver has approximate estimates of the nonlinear functions, $\hat{f}_\text{AM}(\cdot)$ and $\hat{f}_\text{PM}(\cdot)$. In practice, we initialize the procedure with a linear model for the magnitude, $\hat{f}_\text{AM}(x_n)=x_n$, and a distortionless phase, $\hat{f}_\text{PM}(x_n)=0$. In Section \ref{sec:results} we will show that this simple initialization is sufficient to ensure reliable convergence and fast performance in all considered scenarios. The channel length $L$ is assumed to be known at the receiver.

Considering that the nonlinear coefficients are fixed and that $\varepsilon_n$ is linear in $\hat{\mathbf{h}}$, we rewrite problem \eqref{eq:min_mse} as
\begin{align}
    \operatorname*{minimize}_{\hat{\mathbf{h}}}
    \quad
    & \mathbb{E}\left\{
    \left|
    y_n - \hat{\mathbf h}^H\mathbf{u}_n
    \right|^2
    \right\},
    \tag{P2}
    \label{eq:min_taps}
\end{align}
where $\mathbf{u}_n = \mathbf{D}_n \mathbf{C}_n^T \mathbf{f}_{\mathrm{AM}}$ is the effective input signal to the linear channel estimator.

This optimization problem is quadratic and reduces to the standard minimum MSE (MMSE) estimation. The optimal filter coefficients are given by
    \begin{align}
    \hat{\mathbf{h}}^\star = \mathbf{R}_{u}^{-1}\mathbf{r}_{uy}=
    \frac{2}{||\mathbf{f}_{\text{AM}}||^2}\mathbf{r}_{uy},
    \label{eq:optimal_taps}
\end{align}
where $\mathbf{R}_{u} = \mathbb{E}\{\mathbf{u}_n \mathbf{u}_n^H\}$ is the autocorrelation matrix of the effective input signal, and $\mathbf{r}_{uy} = \mathbb{E}\{\mathbf{u}_n y_n^*\}$ is the cross-correlation vector between the effective input and the received signal. 
As shown in Appendix~\ref{app:covariance_u}, the structure induced by the DCT basis leads to $\mathbf{R}_{u}=({||\mathbf{f}_{\text{AM}}||^2}/2)\mathbf{I}$, which only depends on the energy of the magnitude DCT coefficients.

The conventional MMSE solution requires explicit estimation of second-order statistics and a matrix inversion, both of which are costly and must be recomputed whenever the signal statistics vary. In contrast, the DCT-induced isotropy removes the need for matrix inversion entirely, reducing the solution to a simple scaling of the cross-correlation vector. This significantly lowers complexity and improves numerical stability, while avoiding any reliance on eigenvalue decomposition or matrix conditioning.

The same structure also motivates an adaptive implementation. Instead of computing ensemble averages, the channel can be estimated online using a stochastic gradient strategy that updates the solution using the current observation. This leads to a normalized LMS algorithm of the form
\begin{align}
    \hat{\mathbf{h}}^{\{n+1\}}&= \hat{\mathbf{h}}^{\{n\}}-\mu\frac{\partial|\varepsilon_n|^2}{\partial\hat{\mathbf{h}}}
    =\hat{\mathbf{h}}^{\{n\}}
    +\frac{2\alpha}{||\mathbf{u}_n||^2}
    \mathbf{u}_n\varepsilon_n^*\nonumber\\
    &=\hat{\mathbf{h}}^{\{n\}}
    +\frac{4\alpha}{||\mathbf{f}_{\text{AM}}||^2}
    \mathbf{u}_n\varepsilon_n^*,
\end{align}
where $\mu=\alpha/||\mathbf{u}_n||^2 $ is the normalized step-size and $\alpha$ corresponds to the miss-adjustment parameter. The last equality follows from the unitary modulus of $\mathbf{D}_n$ and the diagonal structure of $\mathbf{R}_c$.

A key consequence of the DCT model is that it eliminates direct dependence on the second-order statistics of the original input signal $x_n$, replacing it with a structured and deterministic feature space induced by an orthogonal basis expansion. In this setting, the covariance matrix $\mathbf{R}_u$ becomes diagonal, implying complete statistical decoupling between channel coefficients and yielding an eigenvalue spread equal to one, which provides optimal conditioning and uniform convergence across all channel taps. 

\subsection{Nonlinear Magnitude Distortion}
This section focuses on the estimation of the magnitude DCT coefficients $\mathbf{f}_\text{AM}$. Since $\hat{y}_n$ is scalar-valued, \eqref{eq:rx_estimate} can be rewritten as
\begin{align}
    \hat y_n
    &=
    \mathbf f_\mathrm{AM}^T \mathbf C_n
    \mathbf{D}_n
    \hat{\mathbf h}^*,
    \label{eq:rx_transpose}
\end{align}
which is linear in $\mathbf f_\mathrm{AM}$. Thus, we can rewrite problem \eqref{eq:min_mse} as
\begin{align}
    \operatorname*{minimize}_{{\mathbf{f}}_\text{AM}}
    \quad
    & \mathbb{E}\left\{
    \left|
    y_n - {\mathbf{f}}_\text{AM}^T\mathbf{v}_n
    \right|^2
    \right\}
    \tag{P3}
    \label{eq:min_AM}
\end{align}
where $\mathbf{v}_n = \mathbf C_n \mathbf{D}_n \hat{\mathbf h}^*$ is the effective input signal to the linear estimator. As with problem \eqref{eq:min_taps}, this problem also reduces to the MMSE estimation, with the following optimal filter coefficients:
\begin{align}
    {\mathbf{f}}_\text{AM}^\star &= \Re\{\mathbf{R}_{v}\}^{-1}\,
    \Re\{\mathbf{r}_{vy}\}=\left(\mathbb{E}\left\{||\hat{\mathbf{h}}||^2\mathbf{R}_c\right\}\right)^{-1}
    \Re\{\mathbf{r}_{vy}\}\nonumber\\
    &= \frac{2}{Q||\hat{\mathbf{h}}||^2}\Re\{\mathbf{r}_{vy}\},
    \label{eq:optimal_am}
\end{align}
where $\mathbf{R}_{v} = \mathbb{E}\{\mathbf{v}_n \mathbf{v}_n^H\}$ is the autocorrelation matrix of the effective input signal, and $\mathbf{r}_{vy} = \mathbb{E}\{\mathbf{v}_n y_n^*\}$ is the cross-correlation vector between the effective input and the received signal. A detailed derivation of this result is provided in Appendix \ref{app:a}. 

As in the channel estimation problem, the orthogonality of the DCT basis induces a diagonal covariance matrix. 
Consequently, the estimation of the nonlinear magnitude coefficients also admits an efficient stochastic gradient implementation that avoids explicit covariance estimation and matrix inversion.

The gradient of the instantaneous error with respect to the magnitude coefficients is
\begin{align}
    \frac{\partial|\varepsilon_n|^2}{\partial\mathbf{f}_\text{AM}}
    &=\frac{\partial\varepsilon_n}{\partial\mathbf{f}_\text{AM}}
    \varepsilon_n^*+
    \frac{\partial\varepsilon_n^*}{\partial\mathbf{f}_\text{AM}}
    \varepsilon_n=
    -\mathbf{v}_n\varepsilon_n^*
    -\mathbf{v}_n^*\varepsilon_n
    \nonumber\\
    &=-2\Re\left(\mathbf{v}_n\varepsilon_n^*
    \right),
\end{align}
and the corresponding LMS recursion is given by
\begin{align}
    \mathbf{f}_\text{AM}^{\{n+1\}}&= \mathbf{f}_\text{AM}^{\{n\}}-\mu\frac{\partial|\varepsilon_n|^2}{\partial\mathbf{f}_\text{AM}^{}}
    =\mathbf{f}_\text{AM}^{\{n\}}
    +\frac{2\alpha}{||{\mathbf{v}}_n||^2}
    \Re\left\{\mathbf{v}_n\varepsilon_n^*\right\}
    \nonumber\\
    &=\mathbf{f}_\text{AM}^{\{n\}}
    +\frac{4\alpha}{Q||\hat{\mathbf{h}}||^2}
    \Re\left\{\mathbf{v}_n\varepsilon_n^*\right\}
\end{align}

Therefore, the same benefits observed in the adaptive channel estimation stage are preserved in the estimation of the nonlinear magnitude distortion.

\subsection{Nonlinear Phase Distortion}
This section focuses on the estimation of the phase DCT coefficients $\mathbf{f}_\text{PM}$. Although the phase distortion is linear with respect to the parameter vector, i.e., $\hat{f}_\text{PM}(x_n)=\mathbf{f}_\text{PM}^T\mathbf{c}_n$, the parameters appear inside a complex exponential. Therefore, the MSE in \eqref{eq:min_mse} becomes nonconvex, which prevents the existence of a closed-form global solution. To circumvent this issue, we reformulate the estimation problem in the phase domain, where the dependence on the parameters remains linear. 

We first remove ISI using the current estimates of the channel and nonlinear magnitude distortion, leading to the equalized received signal,
\begin{equation}
    y_n^\text{eq}=y_n-
    \sum_{\ell=1}^{L-1}\hat{h}_\ell^*\hat{f}({x}_{n-\ell})\approx 
    {f}({x}_n)+w_n,
    \label{eq:rx_equalized}
\end{equation}
where the equalization is performed in the time domain. If the channel and nonlinear magnitude estimates are sufficiently accurate, the residual ISI becomes negligible and the remaining perturbation is dominated by AWGN. Under this assumption, the phase of the equalized observation, $\angle\, y_n^\text{eq}$, provides the ML estimate of the underlying signal phase \cite[Chapter 5]{proakis2008digital}.

At sufficiently high signal-to-noise ratio (SNR), $\angle\, y_n^\text{eq}$ admits the first-order approximation:
\begin{align}
    \angle\, y_n^\text{eq}
    &\approx
    \angle\, f(x_n)
    +
    \Im\left\{
    \frac{w_n}{f(x_n)}
    \right\}\nonumber\\
    &=\phi_n+
    f_\text{PM}(|x_{n}|)+
    \Im\left\{
    \frac{w_n}{f(x_n)}
    \right\},
    \label{eq:phase_linear}
\end{align}
where the phase of the equalized received signal decomposes into an additive structure, which enables a direct linear parameterization of the nonlinear phase distortion using the DCT model. 

The receiver models the phase of the equalized signal as
\begin{align}
    \angle\, \hat{y}_n^\text{eq}=\phi_n+
    \hat{f}_\text{PM}(|x_{n}|)=
    \phi_n+\mathbf{f}_\text{PM}^T\mathbf{c}_n,
    \label{eq:phase_hat}
\end{align}
where the phase of the original OFDM sample $\phi_n$ is obtained from the corresponding pilot sample $x_n$.

Defining the estimation error directly over the phase angles is problematic because phase is a circular variable defined modulo $2\pi$. Thus, small angular differences near the phase discontinuities at $\pm\pi$ may produce arbitrarily large Euclidean errors. To avoid these discontinuities, the error is instead defined over the unit circle:
\begin{align}
    \epsilon_n^2&=\Big|\text{e}^{j\angle\, y_n^\text{eq}} -
    \text{e}^{j\angle\, \hat{y}_n^\text{eq}}\Big|^2=
    2-2\cos\left(
    \angle\, y_n^\text{eq}-\angle\, \hat{y}_n^\text{eq}
    \right)
    \label{eq:mse_phase}
\end{align}

The DCT coefficients are then designed to minimize this phase error, which can be rewritten as
\begin{align}
    \operatorname*{maximize}_{{\mathbf{f}_\text{PM}}}
    \quad
    & \mathbb{E}\left\{
    \cos\left(
    \angle\, y_n^\text{eq} -\mathbf{f}_\text{PM}^T\mathbf{c}_n - \phi_n
    \right)
    \right\}
    \tag{P4}
    \label{eq:min_phase}
\end{align}

Problem \eqref{eq:min_phase} is maximized when the phase difference approaches zero, driving the DCT coefficients toward the nonlinear phase distortion. This cost function is commonly encountered in phase synchronization and phase-locked loop (PLL) architectures \cite[Chapter 5]{proakis2008digital}.


Since problem \eqref{eq:min_phase} remains nonconvex and does not admit a closed-form solution, we resort to a stochastic gradient-based procedure to iteratively update the DCT coefficients. The gradient of \eqref{eq:min_phase} with respect to the phase parameters is given by
\begin{align}
    \frac{\partial||\epsilon_n||^2}{\partial\mathbf{f}_\text{PM}}
    &=-2\mathbf{c}_n\sin\left(
    \angle\, y_n^\text{eq}-\angle\, \hat{y}_n^\text{eq}
    \right),
\end{align}
which yields the recursive update
\begin{align}
    \mathbf{f}_{\text{PM}}^{\{n+1\}}&=
    \mathbf{f}_{\text{PM}}^{\{n\}}-\mu\frac{\partial||\epsilon_n||^2}{\partial\mathbf{f}_\text{AM}}
    \nonumber\\
    &=\mathbf{f}_{\text{PM}}^{\{n\}}+\frac{4\alpha}{Q}\mathbf{c}_n\sin\left(
    \angle\, y_n^\text{eq} -\mathbf{f}_\text{PM}^T\mathbf{c}_n - \phi_n
    \right),
\end{align}
where the step-size normalization is chosen with respect to the power of the effective input signal, i.e., $||\mathbf{c}_n||^2=Q/2$.

Although the update is not LMS in the strict MMSE sense, the orthogonal DCT representation preserves the same conditioning advantages, including diagonal covariance structure and uniform convergence behavior.

Finally, it is important to emphasize that this phase-domain reformulation comes at a cost: The extraction of the phase is a nonlinear operation that destroys the additive and Gaussian structure of the noise. Therefore, operating in the phase domain requires sufficiently high SNR conditions in order for \eqref{eq:phase_linear} and the subsequent estimation to remain valid. In the simulation section, we will illustrate that this requirement corresponds to approximately 10 dB of SNR.


\subsection{Joint Estimation Algorithm}
In the previous subsections we derived the optimal estimates of the channel coefficients and the DCT-based nonlinear distortion parameters separately. However, each derivation assumed that the remaining parameter set was known. To address this limitation, we now propose a joint estimation framework.

We consider an alternating optimization strategy summarized in Algorithm~\ref{alg: channel_and_dct}. Given a training sequence $\{\mathbf{x}_n\}$ and the corresponding received samples $\{{y}_n\}$, the receiver iteratively updates (i) the multipath channel coefficients, (ii) the magnitude DCT coefficients, and (iii) the phase DCT coefficients. Each update is performed using a stochastic gradient rule over all $N_\text{samples}$ OFDM samples. Since the gradients are properly normalized, a single step-size parameter $\alpha$ is used for all updates. This alternating procedure is repeated $N_\text{iter}$ iterations until convergence of the MSE.

\begin{figure*}[t]
    \centering
    \includegraphics[width=0.8\linewidth]{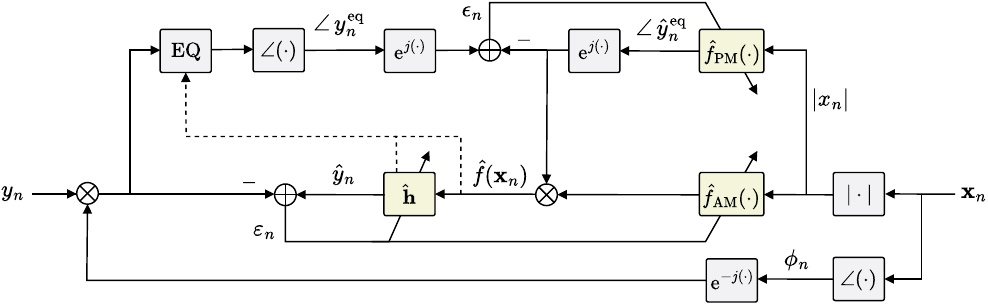}
    \caption{Receiver architecture for estimating the nonlinear frequency-selective channel. Dashed arrows denote auxiliary information provided to the corresponding processing blocks.}
    \label{fig:rx_nonlinear_estimation}
\end{figure*}

Figure~\ref{fig:rx_nonlinear_estimation} illustrates the corresponding receiver architecture for nonlinear frequency-selective channel estimation. The estimated channel and nonlinear distortion parameters are then used in the equalizer (EQ) block. For simplicity, the sample phase $\phi_n$ is subtracted from the received signal $y_n$.

Although not essential to the estimation procedure itself, parameter normalization significantly improves numerical stability and convergence. Specifically, the AM distortion is assumed to be bounded and normalized. Assuming that $f_\text{AM}(\cdot)$ is non-decreasing, the maximum is achieved at the boundary of the domain. For the DCT model, this boundary corresponds to $z=0$, yielding
\begin{equation}
    \hat{f}(0)=\sum_{q=1}^Q F_q
    \cos\left(\frac{\pi(2q-1)}{2N_\text{DCT}}\right)
    \approx \sum_{q=1}^Q F_q,
\end{equation}
where the approximation holds for $N_\text{DCT}\gg Q$. This motivates the normalization of the AM coefficients after each update as
\begin{equation}
    {\mathbf{f}}_\text{AM}\leftarrow \frac{{\mathbf{f}}_\text{AM}}{|\sum_q {{F}}_{\text{AM},q}|}
\end{equation}

\begin{algorithm}[t]
\DontPrintSemicolon
\SetAlgoLined
\KwIn{ $\{\mathbf{x}_n\}$, $\{y_n\}$, $L$, $Q$, $\alpha$, $N_\text{iter}$ }
\vspace*{4 pt}
\KwOut{ $\hat{\mathbf{h}}, \mathbf{f}_\text{AM}, \mathbf{f}_\text{PM}$ }
\vspace*{4 pt}
Initialize $\mathbf{f}_\text{AM},\mathbf{f}_\text{PM}$ randomly (e.g., line coefficients)\;
\vspace*{4pt}
 \For{\upshape $i=1,2,\dots,N_\text{iter}$}{
    \vspace*{4pt}
    \For{\upshape $n=1,2,\dots,N_\text{samples}$}{
        \vspace*{4pt}
        $\mathbf{u}_n=\mathbf{D}_n\mathbf{C}_n^T\,{\mathbf{f}_\text{AM}}$ \;
        \vspace*{4pt}
        $\varepsilon_n=y_n-\hat{\mathbf{h}}^H\mathbf{u}_n$ \;
        \vspace*{4pt}
        $\hat{\mathbf{h}}^{\{n+1\}}=\hat{\mathbf{h}}^{\{n\}}
        +\frac{4\alpha}{||\mathbf{f}_{\text{AM}}||^2}
        \mathbf{u}_n\varepsilon_n^*$ \;
    }
    \vspace*{4pt}
    \For{\upshape $n=1,2,\dots,N_\text{samples}$}{
        \vspace*{4pt}
        $\mathbf{v}_n=\mathbf{C}_n\mathbf{D}_n\,{\hat{\mathbf{h}}^*}$ \;
        \vspace*{4pt}
        $\varepsilon_n=y_n-{\mathbf{f}_\text{AM}}^T\mathbf{v}_n$ \;
        \vspace*{4pt}
        $\mathbf{f}_\text{AM}^{\{n+1\}}=\mathbf{f}_\text{AM}^{\{n\}}
        +\frac{4\alpha}{Q||\mathbf{f}_{\text{AM}}||^2}\,
        \Re\left\{\mathbf{v}_n\varepsilon_n^*\right\}$ \;
    }
    \vspace*{4pt}
    \For{\upshape $n=1,2,\dots,N_\text{samples}$}{
        \vspace*{4pt}
        $y_n^\text{eq}=y_n-\sum_{\ell=1}^{L-1}\hat{h}_\ell^*\hat{f}({x}_{n-\ell})$ \;
        \vspace*{4pt}
        $\epsilon_n=\angle\,y_n^\text{eq}-\left({\mathbf{f}_\text{PM}}^T\mathbf{c}_n+\phi_n\right)$ \;
        \vspace*{4pt}
        $\mathbf{f}_\text{PM}^{\{n+1\}}=\mathbf{f}_\text{PM}^{\{n\}}
        +\frac{4\alpha}{Q}\,
        \mathbf{c}_n\sin\left(\epsilon_n\right)$ \;
    }
 }
 \caption{Joint estimation of the nonlinear frequency-selective channel}
 \label{alg: channel_and_dct}
\end{algorithm}

\subsection{Computational Complexity}
Algorithm \ref{alg: channel_and_dct} consists of three main blocks:
\begin{itemize}
    \item \textit{Adaptation of} $\hat{\mathbf{h}}$: 
    The computation of $\mathbf{u}_n$ involves the product of a matrix with a vector and a diagonal matrix, which requires $\mathcal{O}(Q^2)$ operations. The error signal $\varepsilon_n$ requires a length-$L$ inner product, contributing $\mathcal{O}(L)$ operations. Finally, the update of $\hat{\mathbf{h}}$ involves computing a norm, a vector scaling and addition of length $L$, resulting in $\mathcal{O}(Q+L)$ operations. Therefore, the complexity of the linear channel adaptation stage is $\mathcal{O}(Q^2 + L)$ per sample.
    
    \item \textit{Adaptation of} $\mathbf{f}_\text{AM}$: The computation of $\mathbf{v}_n$ requires two vector products, which incurs $\mathcal{O}(QL)$ operations. The error computation involves a length-$Q$ inner product, resulting in $\mathcal{O}(Q)$ complexity, while the LMS update consists of a vector scaling and addition of length $Q$, also $\mathcal{O}(Q)$. Hence, the overall complexity of the $\mathbf{f}_\text{AM}$ adaptation step is $\mathcal{O}(QL)$ per sample.
    
    \item \textit{Adaptation of} $\mathbf{f}_\text{PM}$\textbf{}: The equalization step requires $\mathcal{O}(L)$ operations and the computation of the error term $\epsilon_n$ requires $\mathcal{O}(Q)$ operations per sample. The LMS update 
    scales linearly with $Q$, leading to $\mathcal{O}(Q)$ operations. Therefore, the $\mathbf{f}_\text{PM}$ adaptation stage has total complexity of $\mathcal{O}(L + Q)$ per sample.
\end{itemize}

Since these three blocks are executed within nested loops, the overall computational complexity of the algorithm is $\mathcal{O}\left(N_\text{iter} N_\text{samples}(Q^2+QL)\right)$.

This result underscores the relevance of the proposed nonlinear model: Thanks to its high representational efficiency, the DCT can accurately capture nonlinear responses using only a limited number of coefficients (e.g., $Q=6$), substantially reducing the computational cost. In addition, the proposed DCT-based model naturally leads to a simple LMS implementation that avoids matrix inversions, further improving computational efficiency.

\section{Nonlinear Compensation Techniques}
\label{sec:compensation}

ML decoding is computationally infeasible because the number of possible transmitted symbols grows exponentially with the number of subcarriers, making exhaustive search prohibitive. Therefore, OFDM systems instead rely on transmitter or receiver-side compensation techniques to mitigate distortions. This section revisits nonlinear compensation methods for OFDM signals and proposes their integration with the channel estimation method presented in this work.

\subsection{Predistortion}
\label{sec:predistortion}

Predistortion is a transmitter-side technique aimed at linearizing nonlinear distortions \cite{PA_modeling}. The predistorted signal is obtained by independently modifying the magnitude and phase of the OFDM signal $x_n$ through two nonlinear functions, $g_{\mathrm{AM}}(\cdot)$ and $g_{\mathrm{PM}}(\cdot)$. These functions are designed such that the cascade of predistorter and nonlinearity approximates an identity mapping:
\begin{align}
    g(f(x_n))
    =g_\text{AM}\left(f_\text{AM}(|x_n|)\right)
    \text{e}^{j\left(
    \phi_n+f_\text{PM}(|x_n|)
    +g_\text{PM}(|x_n|)
    \right)}
\end{align}
From this expression, the conditions for perfect linearization, $g(f(x_n))=x_n$, are
\begin{align}
    g_\text{AM}(\cdot) =f_\text{AM}^{-1}(\cdot)\quad\quad
    g_\text{PM}(\cdot) =-f_\text{PM}(\cdot)
\end{align}

Hence, phase predistortion reduces to a direct cancellation of the AM–PM distortion, whereas magnitude predistortion requires the computation of the inverse AM–AM characteristic, which constitutes an additional nonlinear regression problem.

\begin{figure}[t]
    \centering

    \begin{subfigure}{\columnwidth}
        \centering
        \includegraphics[width=0.72\linewidth]{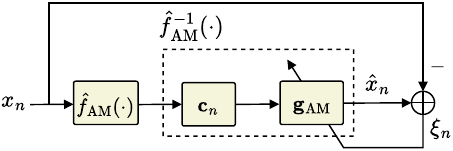}
        \caption{Proposed DCT-based self-supervised learning framework to learn the inverse predistortion function $\hat{f}_\text{AM}^{-1}(\cdot)$ given $\hat{f}_\text{AM}(\cdot)$.}
        \label{fig:predistortion_learning}
    \end{subfigure}

    \vspace{1.em}

    \begin{subfigure}{\columnwidth}
        \centering
        \includegraphics[width=0.9\linewidth]{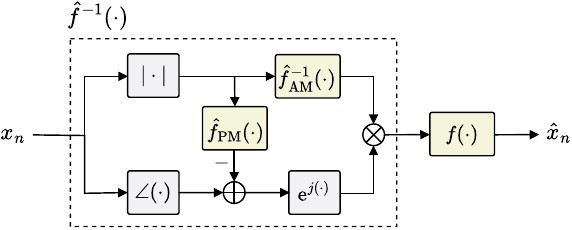} 
        \caption{Predistortion architecture implementing the learned inverse model.}
        \label{fig:predistortion_inference}
    \end{subfigure}

    \caption{Proposed learning-based predistortion framework.}
    \label{fig:predistortion}
\end{figure}

To estimate this inverse, we propose a self-supervised learning framework based on the DCT model, as illustrated in Figure \ref{fig:predistortion}(\subref{fig:predistortion_learning}): A reference signal $x_n$, which may coincide with the training signal used for channel identification, is processed through the estimated magnitude distortion $\hat{f}_\text{AM}(\cdot)$ and subsequently through a parametric inverse model $\hat{f}_\text{AM}^{-1}(\cdot)$. The resulting output is
\begin{equation}
    \hat{x}_n=\hat{f}_\text{AM}^{-1}
    \left(\hat{f}_\text{AM}(x_n)\right),
\end{equation}
which is directly compared to the input sequence $x_n$, yielding a self-supervised error signal.

When the inverse is represented using a DCT model, the estimate can be written as $\hat{x}_n = \mathbf{g}_\text{AM}^T \mathbf{c}_n$, where $\mathbf{g}_\text{AM} \in \mathbb{R}^Q$ contains the DCT coefficients of the inverse magnitude distortion. The error signal is therefore
\begin{equation}
    \xi_n=x_n-\hat{x}_n=
    x_n-\mathbf{g}_\text{AM}^T\,\mathbf{c}_n,
    \label{eq:error_inv}
\end{equation}
which is linear in the coefficients. Once again, this structure enables efficient adaptation using the LMS algorithm:
\begin{equation}
    \mathbf{g}_\text{AM}^{\{n+1\}}= \mathbf{g}_\text{AM}^{\{n\}}+\frac{4\alpha}{Q}\mathbf{c}_n\xi_n
    \label{eq:inv_am}
\end{equation}

Regarding the phase predistortion, the receiver computes the corresponding DCT coefficients as $\mathbf{g}_\text{PM}=-\mathbf{f}_\text{PM}$. Figure \ref{fig:predistortion}(\subref{fig:predistortion_inference}) illustrates the deployment of the proposed predistortion framework, where both magnitude and phase distortions are compensated once the inverse functions have been learned. As a result, the transmitted OFDM signal is effectively linearized, enabling conventional one-tap frequency-domain channel equalization at the receiver.

This formulation preserves the advantages of the DCT model: decorrelated coefficients, bounded gradients and stable convergence behavior. However, since the inverse function is generally more complex than the forward nonlinear characteristic, a larger number of coefficients $Q$ might be required. This only increases convergence time proportionally, which can be theoretically bounded by
\begin{equation}
    T_\kappa = \frac{Q\ln(1/\kappa)}{2\alpha},
    \label{eq:time_convergence}
\end{equation}
where $\kappa$ indicates convergence when the residual error in the coefficients has reduced
to $\kappa$ \cite{dct_apn}. Furthermore, the self-supervised learning process is fully local (i.e., at the transmitter or receiver only) and operates under idealized conditions with infinite SNR, such that performance is determined solely by the representational capacity of the DCT model, i.e., required number of coefficients.

The computational complexity of estimating the inverse function using the DCT-based model scales linearly with both the number of training samples and the number of coefficients, yielding an overall complexity of $\mathcal{O}(N_{\text{samples}}Q)$. Similarly, the computational complexity of the predistortion stage is dominated by the evaluation of the DCT-based model, which scales with $Q$. Therefore, the overall predistortion complexity per sample is $\mathcal{O}(Q)$.

An important remark is that directly estimating the inverse function at the receiver is not advisable. Although such an approach may appear to simplify the processing chain, it violates the ML structure of the estimation problem and leads to degraded performance even at high SNR. This degradation arises because nonlinear inversion at the receiver transforms additive Gaussian noise into a signal-dependent and non-Gaussian process, significantly complicating estimation and potentially amplifying noise. For this reason, the preferred approach is to first estimate the forward nonlinear characteristic $\hat{f}_{\mathrm{AM}}(\cdot)$ and subsequently derive its inverse in a controlled manner, as proposed. We refer the reader to \cite{nature_takesover} for a comprehensive theoretical comparison and evaluation of both approaches.

Finally, the proposed predistortion scheme differs from conventional approaches in two key aspects. First, it employs a DCT-based model, which significantly reduces the complexity of estimation. Second, the inverse estimation might be performed at the receiver. In traditional digital predistortion architectures, adaptation is carried out at the transmitter, which requires additional hardware to capture the output signal, perform analog-to-digital conversion, and ensure precise synchronization between transmitted and feedback signals. These requirements become increasingly demanding as communication bandwidths grow, leading to higher sampling frequencies and increased power consumption. In contrast, the proposed approach leverages the existing receiver processing chain, thereby avoiding dedicated feedback hardware and benefiting from already available digital signal processing resources.

\subsection{Postdistortion}


Postdistortion techniques are receiver-side compensation methods that aim to correct nonlinear distortion after it has already affected the transmitted signal \cite{post-distortion}.

The frequency-domain received signal at subcarrier $k$ can be written as
\begin{align}
    Y_k=\mathcal{F}\left(y_n\right)
    &=H_k\,\mathcal{F}\left(f(x_n)\right)+W_k\nonumber\\
    &=H_kG_{k,k}X_k + 
    \underbrace{H_k\sum_{m\neq k}G_{k,m}X_m}_{\text{ICI}} + W_k,
\end{align}
where $H_k$ denotes the linear channel frequency response, $W_k$ is the additive noise in the frequency domain and $G_{k,m}$ represents the nonlinear coupling coefficients induced by $f(\cdot)$, mapping subcarrier $m$ onto subcarrier $k$. The off-diagonal capture the ICI introduced by the nonlinear transmitter.

Despite the presence of ICI, the linear multipath channel still appears as a multiplicative per-subcarrier factor $H_k$. This preserves the applicability of conventional one-tap frequency-domain equalization. Given a channel estimate $\hat{H}_k = \mathcal{F}({\hat{h}_n})$, the equalized signal becomes
\begin{equation}
    Y_k^\text{eq}=\frac{Y_k}{\hat{H}_k}\approx
    G_{k,k}X_k + 
    \sum_{m\neq k}G_{k,m}X_m+ \frac{W_k}{H_k}
    \label{eq:rx_eq}
\end{equation}

This shows that channel equalization effectively removes the linear multipath effect, while nonlinear distortion remains embedded as ICI terms. In the time domain, \eqref{eq:rx_eq} corresponds to
\begin{equation}
    y_n^\text{eq}=\mathcal{F}^{-1}(Y_k^\text{eq})\approx
    f(x_n)+\tilde{w},
\end{equation}
which highlights that the linear and nonlinear distortion can be separated under the assumption of a known channel response.

A large class of compensation techniques builds on the assumption that an accurate channel estimate $\hat{\mathbf{h}}$ is available at the receiver. These include:
\begin{itemize}
    \item \textit{Iterative decoding} (\cite{cioffi_iterative, gregorio}): reconstructs the distortion at the receiver using a known model of the distortion and detected data symbols, and successively refines and subtracts this estimated distortion from the received signal to improve symbol estimation. It operates in the frequency domain and assumes perfect knowledge of both the nonlinear response and the frequency-selective channel for quasi-ML detection.

    \item \textit{Neural network}
    (\cite{nn_heavy}): learns the inverse of the nonlinear channel, enabling direct reconstruction of the undistorted signal without requiring an explicit analytical model. The frequency-selective channel response is provided as input.
    
    \item \textit{Decision-aided reconstruction}
    (\cite{clipping_dar}): restores the clipped signal using ML detection to recover regrown peaks, replacing clipped samples with reconstructed values and repeating the process until convergence. It assumes one-tap frequency-domain equalization is available to compensate for multipath propagation.
    
    \item \textit{Clipping noise cancellation}
    (\cite{clipping_dar2, clipping_dar3}): estimates the clipping distortion by regenerating and re-clipping a time-domain signal constructed from detected symbols, then subtracts the estimated distortion from the received signal. It assumes one-tap frequency-domain equalization is available to compensate for multipath propagation.

    \item \textit{Compressive sensing}
    (\cite{CS1, CS2}): exploits the sparsity of the clipping distortion in the time domain and reconstructs it from a subset of received subcarriers using sparse recovery methods such as Least Absolute Shrinkage and Selection Operator (LASSO). It assumes one-tap frequency-domain equalization is available to compensate for multipath propagation.
\end{itemize}

While these methods differ in their assumptions and scope (see \cite{post-distortion} for a detailed discussion), they generally rely on the availability of accurate side information, including an estimate of the frequency-selective channel $\hat{\mathbf{h}}$ and, in some cases, an estimate of the nonlinear distortion $\hat{f}(\cdot)$. In practice, however, these quantities are not directly available in nonlinear scenarios with induced ICI. Most existing approaches therefore implicitly assume that such estimates can be obtained, despite the fact that the underlying estimation problem is itself nontrivial. This highlights the need for a dedicated channel estimation framework that can operate under nonlinear conditions, which is precisely the gap addressed in this work.

Among the receiver-side compensation techniques discussed above, iterative decoding plays a central role due to its exploitation of the nonlinear distortion knowledge. These algorithms assume the following decomposition:
\begin{equation}
    f(x_n)=x_n + d_n,
    \label{eq:ofdm_linearization}
\end{equation}
where $d_n$ encapsulates the nonlinear component of the signal, which is treated as noise. This linearization allows to work with OFDM in the frequency domain, as if there was no ICI. First, hard-decoding is applied over each subcarrier to obtain an estimate of the transmitted digital symbol, $\hat{X}_k=\langle Y_k/\hat{H}_k\rangle$. Then, the additive distortion $d_n$ is estimated in the time domain and used to correct the decoding in an iterative fashion. 


These decision-feedback algorithms assume perfect knowledge of both the nonlinear response and the frequency-selective channel, which can be provided by the proposed estimation algorithm. 



\section{Simulation Results}
\label{sec:results}

This section focuses on a satellite communication use case, where nonlinear distortions introduced by high-power amplifiers (HPAs) constitute a fundamental performance bottleneck \cite{HPA_sat}. In order to maintain power efficiency, satellite transceivers are typically operated close to saturation, which exacerbates nonlinear amplitude and phase distortions. While classical mitigation strategies based on input back-off can alleviate these impairments, they compromise power efficiency, making them poorly suited for modern non-terrestrial networks where energy are particularly stringent. This challenge is further exacerbated in OFDM systems, which have historically been regarded as problematic in space applications due to their inherently high PAPR, which necessitates large back-off margins and thus further reduces the effective utilization of onboard amplifier power.

Within this context, we first evaluate the proposed algorithm for nonlinear channel estimation. We then use the resulting channel estimates within two compensation schemes to study their effectiveness in mitigating nonlinear effects at the transmitter and receiver, respectively.

\begin{figure*}[t]
    \centering

    



    \begin{subfigure}[t]{\textwidth}

    \begin{subfigure}[b]{0.24\textwidth}
        \centering
        {\footnotesize $\mathrm{MSE}=1.3\times10^{-9}$\par}
        \vspace{0.2em}
        \includegraphics[width=\textwidth]{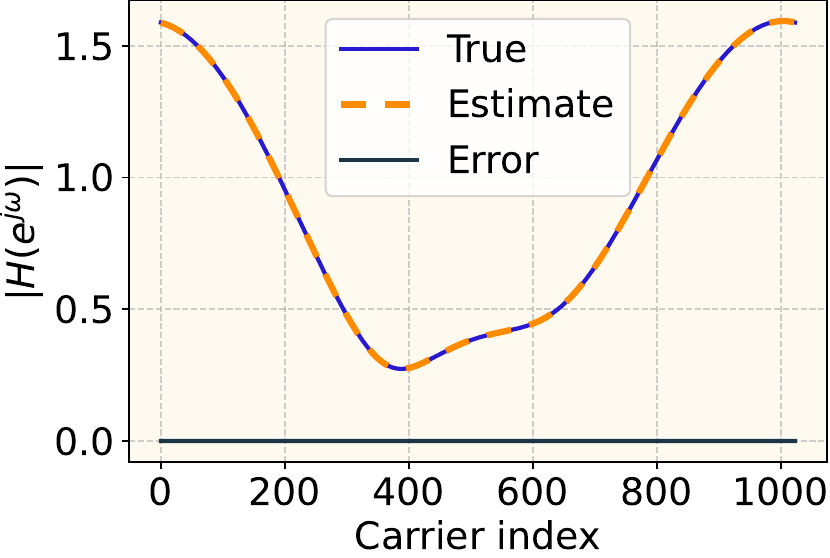}
    \end{subfigure}
    \hfill
    \begin{subfigure}[b]{0.24\textwidth}
        \centering
        \vspace{0.2em}
        \includegraphics[width=\textwidth]{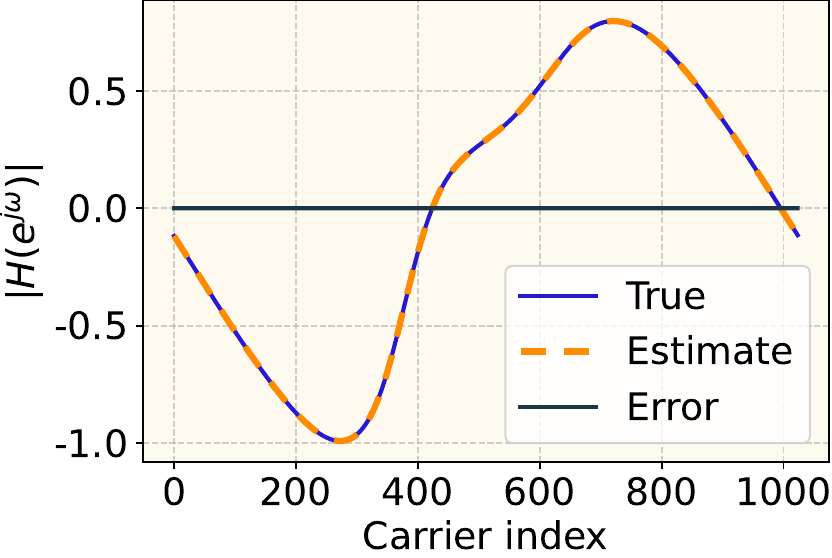}
    \end{subfigure}
    \hfill
    \begin{subfigure}[b]{0.24\textwidth}
        \centering
        {\footnotesize $\mathrm{MSE}=3.0\times10^{-6}$\par}
        \vspace{0.2em}
        \includegraphics[width=\textwidth]{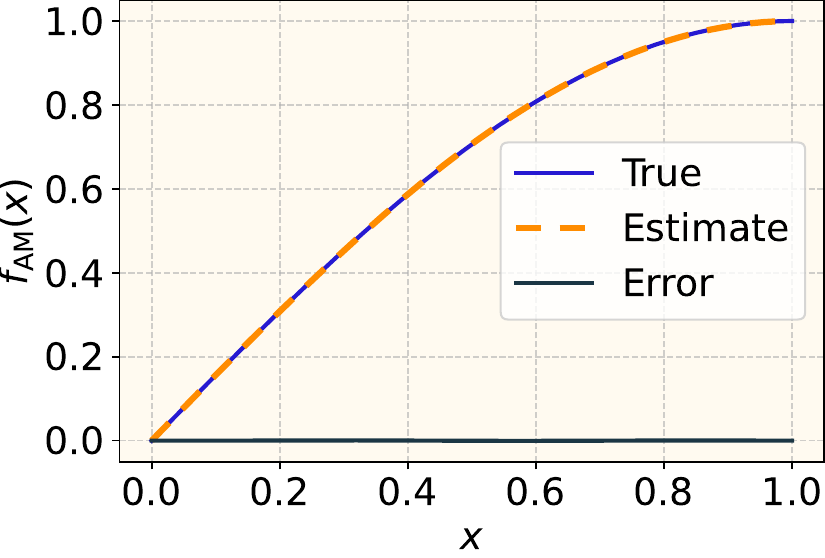}
    \end{subfigure}
    \hfill
    \begin{subfigure}[b]{0.24\textwidth}
        \centering
        {\footnotesize $\mathrm{MSE}=9.0\times10^{-5}$\par}
        \vspace{0.2em}
        \includegraphics[width=\textwidth]{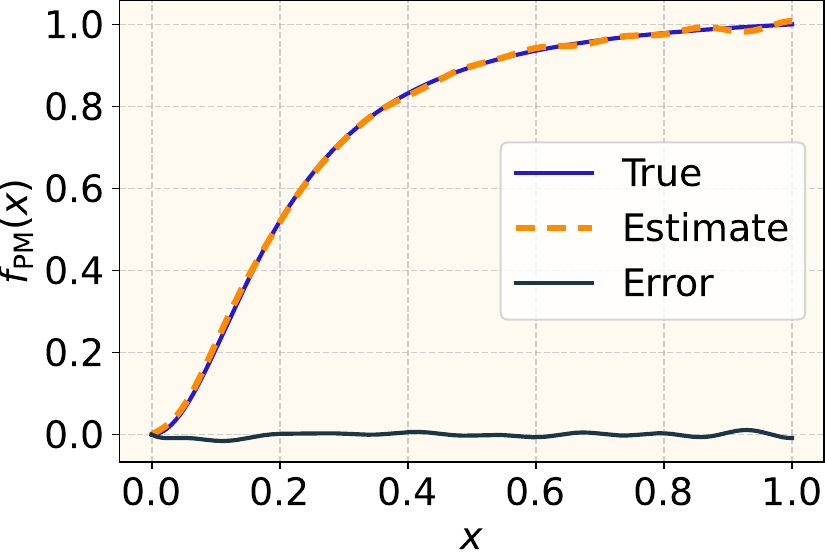}
    \end{subfigure}

    \subcaption{$\text{SNR}^\text{pre}=30$ dB.}
    \label{fig:iir_mdir_dct_snr_10}
    \end{subfigure}

    \vspace{1em}

    \begin{subfigure}[t]{\textwidth}

    \begin{subfigure}[b]{0.24\textwidth}
        \centering
        {\footnotesize $\mathrm{MSE}=1.9\times10^{-6}$\par}
        \vspace{0.2em}
        \includegraphics[width=\textwidth]{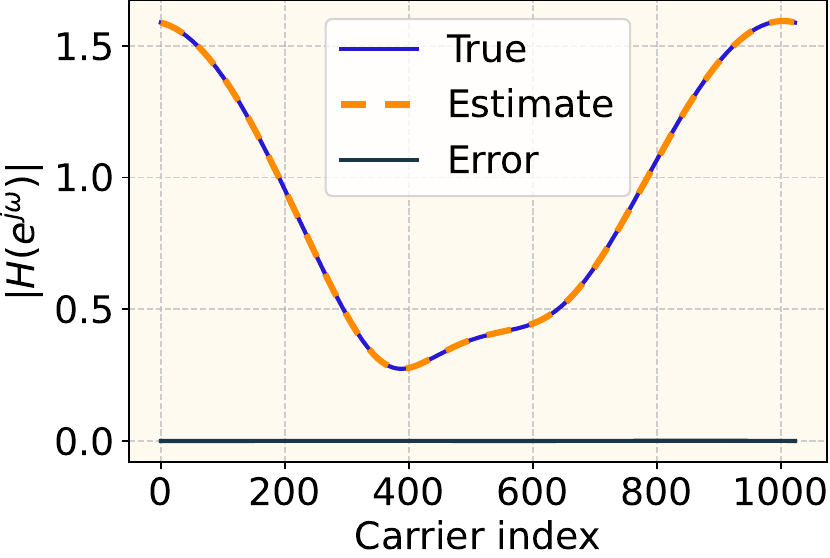}
    \end{subfigure}
    \hfill
    \begin{subfigure}[b]{0.24\textwidth}
        \centering
        \vspace{0.2em}
        \includegraphics[width=\textwidth]{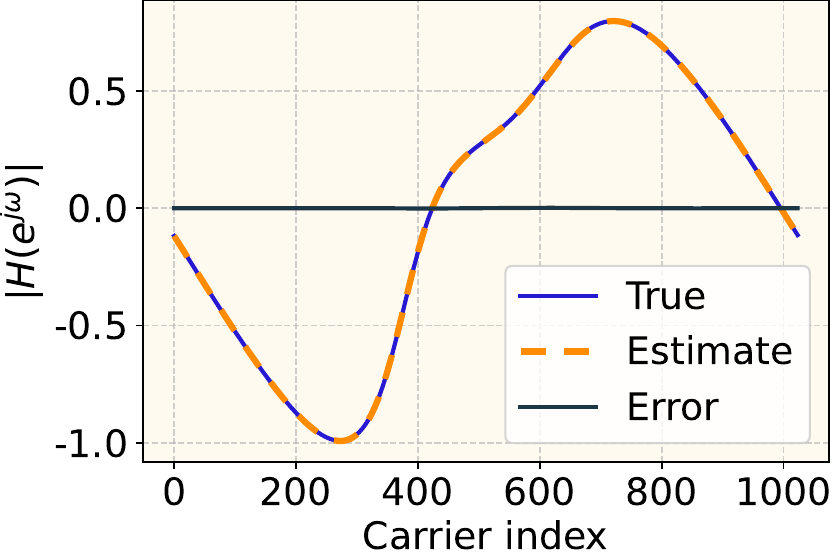}
    \end{subfigure}
    \hfill
    \begin{subfigure}[b]{0.24\textwidth}
        \centering
        {\footnotesize $\mathrm{MSE}=5.0\times10^{-5}$\par}
        \vspace{0.2em}
        \includegraphics[width=\textwidth]{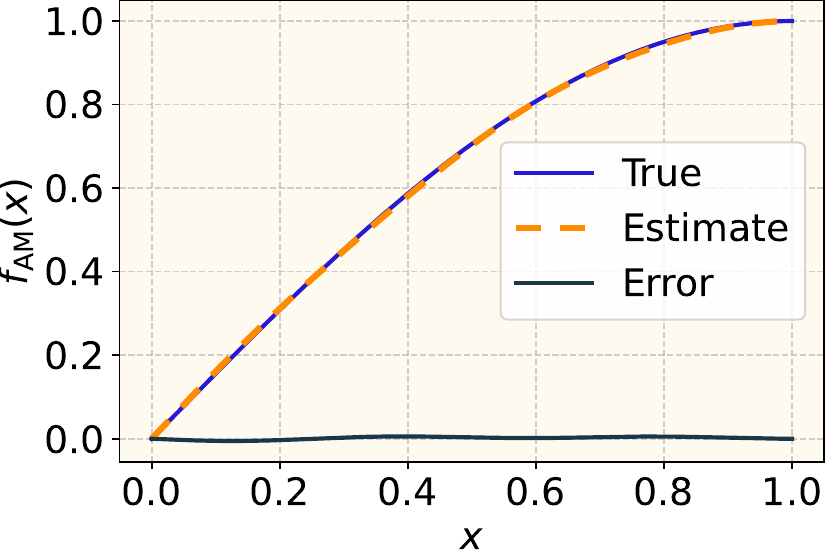}
    \end{subfigure}
    \hfill
    \begin{subfigure}[b]{0.24\textwidth}
        \centering
        {\footnotesize $\mathrm{MSE}=5.7\times10^{-4}$\par}
        \vspace{0.2em}
        \includegraphics[width=\textwidth]{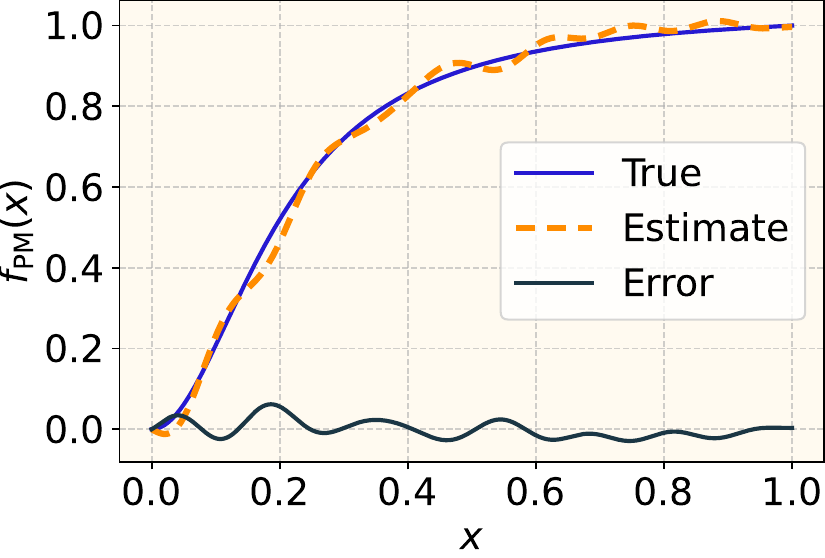}
    \end{subfigure}

    \subcaption{$\text{SNR}^\text{pre}=15$ dB.}
    \label{fig:fir_mdir_dct_snr_10}
    \end{subfigure}

    \caption{Nonlinear channel estimation for different noise configurations. From left to right, the columns show the magnitude response of the frequency-selective channel, the phase response of the frequency-selective channel, the nonlinear amplitude distortion and the nonlinear phase distortion. For each estimate, the corresponding MSE is indicated.}
    \label{fig:mdir_dct}
\end{figure*}

\subsection{Experimental setup}
The simulations are performed using an OFDM system with $N_\text{DFT}=1024$ subcarriers, 16-quadrature amplitude modulation (16-QAM) on each symbol and a cyclic prefix of length 16. The multipath channel has length $L=3$. At the receiver, an overestimated channel length of $\hat{L}=6$ is assumed, which might be truncated afterwards. 

Algorithm \ref{alg: channel_and_dct} is configured with $\alpha=0.1$ and $N_\text{iter}=5$ iterations. A total of $N_\text{samples}=25.000$ samples are transmitted, corresponding to 24 OFDM symbols or 2 transmission time intervals (TTI). For the DCT model, a resolution of $N_\text{DCT}=512$ samples is used, with $Q=6$ coefficients for the amplitude distortion and $Q=12$ for the phase distortion. 

We consider two HPA models: a traveling-wave tube amplifier (TWTA) and a solid-state power amplifier (SSPA). For the TWTA we adopt the Saleh model \cite{saleh} to characterize its nonlinear behavior, given by
\begin{equation}
    f_\text{AM}(x)=\frac{\alpha_\text{AM}|x|}{1+\beta_\text{AM}|x|^{2}}\,,\quad
    f_\text{PM}(x)=\frac{\alpha_\text{PM}|x|^2}{1+\beta_\text{PM}|x|^{2}},
\end{equation}
with parameters $\alpha_\text{AM}=1.7$, $\beta_\text{AM}=0.7$, $\alpha_\text{PM}=26$ and $\beta_\text{PM}=25$. For the SSPA, which exhibits only amplitude distortions, we retain $f_\text{AM}(x)$ and set $f_\text{PM}(x)=0$.

The proposed DCT model was compared in \cite{nature_takesover} against classical polynomial-based representations (e.g., Volterra series) as well as neural network-based models. The results showed that, even in an additive white Gaussian noise (AWGN) scenario with an ideal channel, the DCT-based model achieves substantially lower MSE and significantly faster convergence across all tested configurations. For this reason, in the present work we do not include further comparisons with alternative nonlinear representations, as they have been shown to be less effective and are not expected to outperform the proposed approach.

Finally, we assume that the estimation and subsequent communication (i.e., compensation) happen within the same coherence block, so that the nonlinear channel response does not change. Otherwise, the distortions are static and the selective-fading can be re-estimated with conventional pilot-assisted techniques.

\subsection{Nonlinear Channel Estimation}

To assess the robustness of the proposed estimation scheme, we define the pre-detection SNR at the receiver input as
\begin{equation}
    \text{SNR}^\text{pre}=\frac{\mathbb{E}\Big\{\big|\mathbf{h}^Hf(\mathbf{x}_n)\big|^2\Big\}}{\sigma^2}
\end{equation}

We first address the more challenging case of the TWTA. Figure \ref{fig:mdir_dct} shows the estimated nonlinear channel components together with their corresponding MSE, normalized by the power of the true responses. Figure \ref{fig:mdir_dct}(\subref{fig:iir_mdir_dct_snr_10}) corresponds to the estimation at 30 dB, where all channel components are recovered with negligible error. The phase nonlinearity exhibits the largest estimation error, as justified in the theoretical analysis. 
The proposed method continues to provide accurate estimates down to approximately 15 dB, as illustrated Figure \ref{fig:mdir_dct}(\subref{fig:fir_mdir_dct_snr_10}), which can be regarded as a practical lower operating limit. This requirement is generally not restrictive, since OFDM systems are typically operated at higher SNR values.

\begin{figure}[t]
    \centering
    \includegraphics[width=0.7\columnwidth]{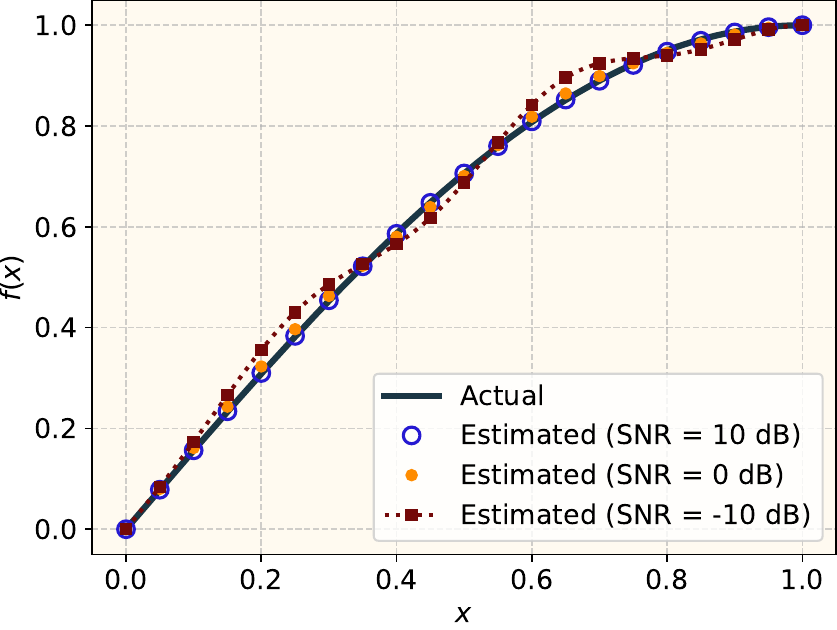}
    \caption{SSPA magnitude distortion and estimated response at different SNR levels.}
    \label{fig:sspa}
\end{figure}

Figure \ref{fig:sspa} presents the results for the SSPA case. We only illustrate the nonlinear response, since the underlying multipath response is perfectly estimated across all considered SNR regimes. Not having to estimate the phase distortion simplifies Algorithm \ref{alg: channel_and_dct}, leading to two key advantages: first, reliable channel estimation is achieved down to $\text{SNR}=0$ dB, and even at $\text{SNR}=-10$ dB the estimation errors remain small enough to ensure excellent nonlinear compensation performance. Second, the proposed approach is highly efficient in terms of training overhead, requiring only 2 OFDM symbols (i.e., 2048 time samples) to accurately capture the nonlinear frequency-selective channel, compared to 24 OFDM symbols in the general setup.

The robustness of the proposed estimation framework is further assessed with respect to the initialization, convergence behavior and assumed channel length. Although the underlying optimization problem is non-convex and the alternate optimization strategy cannot guarantee convergence to the global optimum, the proposed algorithm consistently converges to a local minimum that accurately identifies the channel components under all the considered SNR conditions and nonlinear configurations. This behavior is largely enabled by the adopted initialization, which represents the most general case by assuming an initially linear channel with no nonlinear distortion. Regarding the channel length, underestimating the number of multipath components prevents accurate channel recovery, whereas overestimating it is a safer strategy since redundant components can be removed afterwards. In all simulations, the proposed estimator automatically suppresses the additional multipath components by driving their corresponding coefficients to zero, without introducing artificial channel paths. This behavior is consistent with the ML formulation adopted in this work, where the receiver jointly estimates the complete nonlinear channel response and avoids ambiguous solutions arising from independent fitting of multipath and nonlinear distortion components.


\subsection{Nonlinear Channel Compensation}

Once the nonlinear channel is estimated, we propose two different mitigation strategies:
\begin{itemize}
    \item \textit{Predistortion}: The inverse nonlinear response is represented using $Q=64$ coefficients. The LMS adaptation rule in \eqref{eq:inv_am} is implemented with the same step size $\alpha=0.1$. Training the inverse model requires $2,000$ samples, in agreement with the theoretical prediction in \eqref{eq:time_convergence} for $\kappa=10^{-3}$.

    \item \textit{Iterative-decoding}: The estimated phase distortion is fed back to the transmitter and compensated through predistortion, while the estimated magnitude distortion is used at the receiver to decode the information. The iterative procedure is run for $N_{\text{step}}=5$ iterations.
\end{itemize}

\begin{figure}[t]
    \centering
    \begin{subfigure}{\columnwidth}
        \centering
        \includegraphics[width=\linewidth]{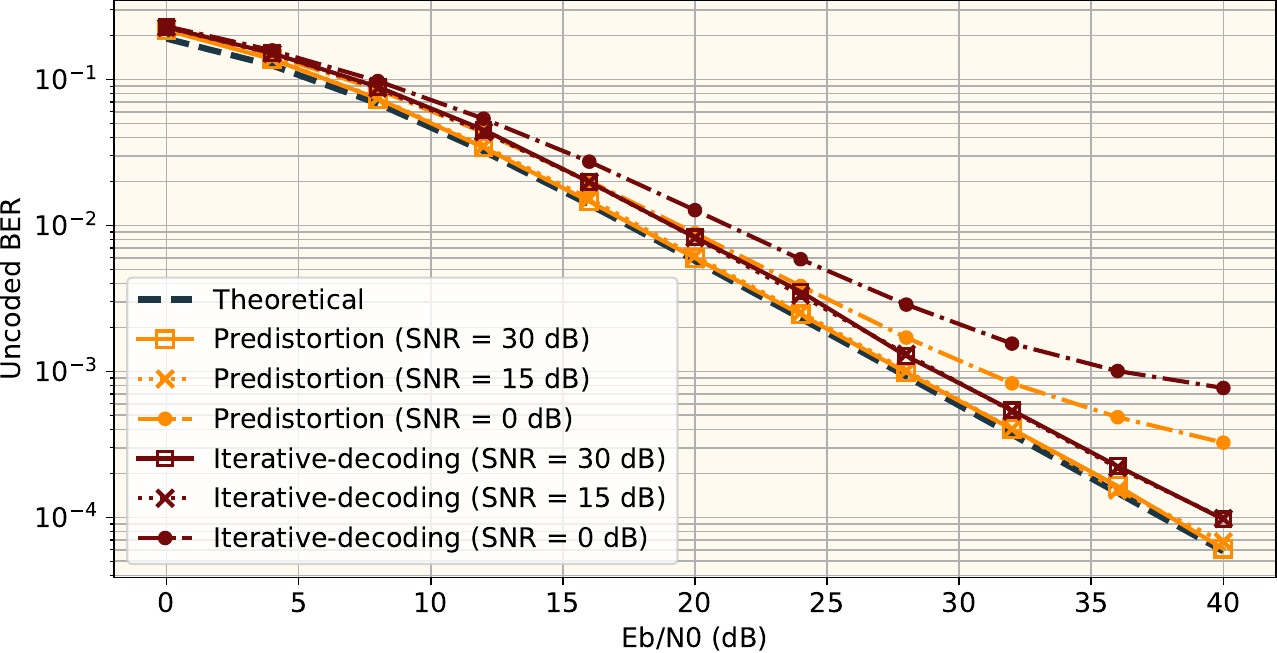}
        \caption{TWTA.}
        \label{fig:twta_ber}
    \end{subfigure}
    
    \vspace{1.em}
    
    \begin{subfigure}{\columnwidth}
        \centering
        \includegraphics[width=\linewidth]{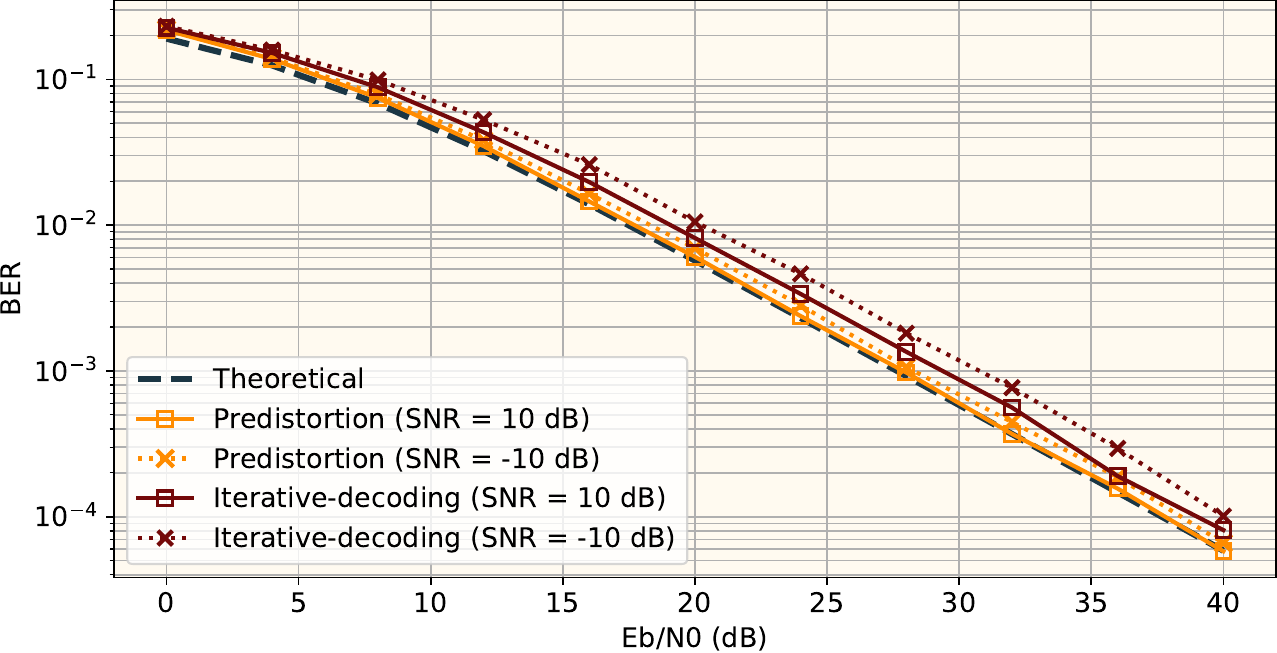}
        \caption{SSPA.}
        \label{fig:sspa_ber}
    \end{subfigure}

    \caption{Comparison of BER performance under different nonlinear distortions, evaluating different compensation schemes with channel estimation performed at different SNR levels.}
    \label{fig:ber}
\end{figure}

Figure \ref{fig:ber} reports the uncoded BER of the OFDM system under different nonlinear distortions and compensation schemes, using channel estimates obtained at various SNR levels. Figure \ref{fig:ber}(\subref{fig:twta_ber}) presents the results for the TWTA model. Although the phase estimate is not fully accurate at 15 dB, as observed in Figure \ref{fig:mdir_dct}(\subref{fig:fir_mdir_dct_snr_10}), the estimated channel is still sufficiently accurate to enable effective compensation with both techniques. Predistortion achieves performance that is nearly indistinguishable from that of an ideal linear OFDM system, owing to the invertibility of the TWTA magnitude response. In contrast, iterative decoding exhibits a small performance degradation, since it relies on a quasi-ML procedure and is therefore inherently suboptimal. When the channel is estimated at 0 dB, an error floor appears at high SNR due to residual estimation inaccuracies. However, up to approximately 30 dB, the performance remains close to that obtained with higher-quality estimates, indicating that the proposed estimator still provides sufficiently accurate channel information for effective nonlinear compensation.

Figure \ref{fig:ber}(\subref{fig:sspa_ber}) shows the SSPA case. In this case, the channel estimation remains reliable even at $-10$ dB. Both compensation schemes exhibit comparable performance, following the same general trend observed in the TWTA case.

Table~\ref{tab:comparison_methods} compares the proposed methods in terms of sample size (i.e., the number of transmitted discrete-time samples required for estimation), sample complexity (i.e., the total number of sample usages across training iterations required to reach a target estimation performance) and model size. Predistortion incurs in a larger model size to represent the inverse magnitude distortion, which does not highly increase the sample complexity because the process is dominated by the channel estimation.

In practice, the choice of technique is likely to be driven by implementation constraints and deployment considerations. In particular, predistortion shifts most of the computational burden to the transmitter, but it relies on the assumption that the magnitude distortion is invertible, which can limit its applicability in strongly nonlinear regimes. In contrast, iterative decoding distributes the processing across both transmitter and receiver. However, this method is more sensitive to severe nonlinearities, even if invertible, under which its performance may degrade or even fail to converge.

\begin{table*}[t]
\centering
\renewcommand{\arraystretch}{1.5}
\caption{Comparison of different nonlinear compensation methods. (TX = Transmitter; RX = Receiver)}

\resizebox{\linewidth}{!}{%
\begin{tabular}{c c c c c c c}
\toprule
\rule{0pt}{3.2ex}
\textbf{Method} &
\textbf{Distortion type} &
\textbf{Deployment} &
\textbf{Sample size} &
\textbf{Sample complexity} &
\textbf{Model size} &
\textbf{Limitations} \\
\midrule

\multirow{2}{*}{\shortstack{DCT-based\\predistortion}}
& AM-AM
& TX
& $2\times10^{3}$
& $3\times10^{4}$
& $73$
& \multirow{2}{*}{\shortstack{Distortions must \\be invertible}} \\
& AM-AM, AM-PM
& TX
& $2\times10^{4}$
& $1\times10^{5}$
& $85$
& \\
\midrule

\multirow{2}{*}{\shortstack{DCT-based\\iterative decoding}}
& AM-AM
& RX
& $2\times10^{3}$
& $1\times10^{5}$
& $9$
& \multirow{2}{*}{\shortstack{Fails under \\severe distortions}} \\
& AM-AM, AM-PM
& TX, RX
& $2\times10^{4}$
& $1\times10^{5}$
& $21$
& \\
\midrule

\shortstack{Neural network\\\cite{nn_heavy}}
& \shortstack{Any\vspace{1.5pt}}
& \shortstack{RX\vspace{3.pt}}
& \shortstack{$2\times10^{9}$\vspace{3.pt}}
& \shortstack{$5\times10^{14}$\vspace{3.pt}}
& \shortstack{$6\times10^{5}$\vspace{3.pt}}
& \shortstack{Depends on\\modulation} \\
\bottomrule
\end{tabular}
}
\label{tab:comparison_methods}
\end{table*}

Table~\ref{tab:comparison_methods} also contrasts the proposed DCT-based methods with neural networks. Neural networks are purely data-driven, operate at the receiver and learn an implicit inverse mapping directly from observations. However, they are structurally less efficient: while the proposed methods rely on a compact set of $10-10^{2}$ parameters, neural networks require on the order of $10^{5}$ parameters, along with substantially higher training and sample complexity. In particular, they require up to $\times 10^{5}$ more training samples and $\times 10^{9}$ more data samples to reach convergence. In contrast, the proposed approach is more transparent, as it explicitly characterizes the channel response through a DCT representation, enabling a physically interpretable decomposition of the distortion. This structure also improves efficiency in two key aspects: (i) it reduces data requirements due to the compact parameterization of the channel, and (ii) it accelerates convergence by exploiting the energy compaction and decorrelation properties of the DCT. Furthermore, unlike neural networks, that are typically trained for specific modulation formats, the proposed methods operate in the time domain and remain independent of the underlying OFDM modulation scheme.

\section{Conclusion}

In this paper, we have proposed a ML framework for estimating nonlinear frequency-selective channels in OFDM communication system. This work addresses the lack of general and computationally efficient estimation procedures capable of jointly handling multipath, nonlinear amplitude and phase distortions in a unified manner. The nonlinear distortions are modeled using a DCT-based representation, which provides a compact and flexible parametrization while leading to a well-conditioned estimation problem with favorable convergence properties. These characteristics enable fast convergence and eliminate the need for step-size tuning or the estimation of second-order statistics.
Numerical results show that the proposed channel estimation can be effectively integrated into different nonlinear compensation strategies, including predistortion and iterative decoding. Across these methods, the resulting system achieves near-ideal BER performance with very limited training overhead. This performance is maintained down to approximately 15 dB SNR in the presence of both amplitude and phase nonlinearities, and down to $-10$ dB when only amplitude distortions are present.
Overall, the proposed approach combines structural efficiency and low computational complexity, enabling fast and robust adaptation, and making it well suited for real-time implementation in multicarrier communication systems under nonlinear channel conditions.

\appendices

\section{}
\label{app:covariance_u}
In this appendix, we derive the autocorrelation matrix $\mathbf{R}_u$. The $(i,j)$th entry of $\mathbf{R}_u$ can be written as
\begin{align}
    \left[\mathbf{R}_u\right]_{i,j}&=
    \mathbb{E}\left\{
    \mathbf{f}_\text{AM}^T\mathbf{c}_{n-i}
    \mathbf{c}_{n-j}^T\mathbf{f}_\text{AM} 
    \,\text{e}^{j(\psi_i-\psi_j)}
    \right\}\nonumber\\
    &=\mathbf{f}_\text{AM}^T\mathbb{E}\left\{
    \mathbf{c}_{n-i}
    \mathbf{c}_{n-j}^T\right\}
    \mathbf{f}_\text{AM}
    \mathbb{E}\left\{
    \text{e}^{j(\psi_i-\psi_j)}
    \right\},
\end{align}
where $\psi_i=\phi_{n-i}+ \mathbf{f}_\text{PM}^T\mathbf{c}_{n-i}$. The second equality follows from the independence assumption between the phase and amplitude processes in an OFDM signal.

Regarding the phase, $\phi_n$ is usually modeled as i.i.d. random variable with distribution $\phi_n\sim\mathcal{U}(0,2\pi)$. Thus, we obtain 
\begin{align}
    \mathbb{E}\left\{
    \text{e}^{j(\psi_i-\psi_j)}
    \right\} =
    \begin{cases}
    1 & \text{if } i=j\\
    0 & \text{if } i\neq j \\
    \end{cases}
\end{align}
and off-diagonal entries reduce to $\left[\mathbf{R}_u\right]_{i,j}=0$. For diagonal entries, we have:
\begin{align}
    \left[\mathbf{R}_u\right]_{i,i}=
    \mathbf{f}_\text{AM}^T
    \mathbb{E}\left\{
    \mathbf{c}_{n-i}
    \mathbf{c}_{n-i}^T
    \right\}
    \mathbf{f}_\text{AM}=
    \mathbf{f}_\text{AM}^T
    \mathbf{R}_c
    \mathbf{f}_\text{AM}=
    \frac{||\mathbf{f}_\text{AM}||^2}{2},
\end{align}
which follows from the orthogonality of the DCT basis and the assumption of independent realizations across time.
$\hfill\square$

\section{}
\label{app:a}
In this appendix we derive the optimal solution of \eqref{eq:min_AM}. First, we expand the cost function:
\begin{equation}
    \mathbb{E}\left\{
    |\varepsilon_n|^2
    \right\}=
    \mathbb{E}\left\{
    |y_n|^2
    \right\}-
    2\Re\left\{
    \mathbf{f}_\text{AM}^T\mathbf{r}_{vy}
    \right\}+
    \mathbf{f}_\text{AM}^T
    \mathbf{R}_{v}
    \mathbf{f_\text{AM}},
\end{equation}
where $\mathbf{R}_{v} = \mathbb{E}\{\mathbf{v}_n \mathbf{v}_n^H\}$ is the autocorrelation matrix of the effective input signal, and $\mathbf{r}_{vy} = \mathbb{E}\{\mathbf{v}_n y_n^*\}$ is the cross-correlation vector between the effective input and the received signal.

Since the objective function of \eqref{eq:min_AM} is quadratic in $\mathbf{f}_\text{AM}$ and the problem in unconstrained, the optimal solution is obtained by setting the gradient of the cost function to zero:
\begin{align}
    \frac{\partial\mathbb{E}\{|\varepsilon_n|^2\}}{\partial\mathbf{f}_\text{AM}}
    &=-2\Re\{\mathbf{r}_{vy}\}+
    \left(\mathbf{R}_{v}+\mathbf{R}_{v}^H\right)\mathbf{f}_\text{AM}\nonumber\\
    &=-2\Re\{\mathbf{r}_{vy}\}+
    2\Re\{\mathbf{R}_{v}\}\mathbf{f}_\text{AM}
\end{align}

Setting the gradient equal to zero yields the optimal solution in \eqref{eq:optimal_am}.

The analytical expression of $\mathbf{R}_v$ follows directly from the same arguments used in the derivation of $\mathbf{R}_u$ in Appendix \ref{app:covariance_u}.
$\hfill\square$

\bibliographystyle{IEEEbib}
\bibliography{refs}

\end{document}